\documentclass[]{aiaa-tc}

\usepackage{amsmath,amssymb,stmaryrd,array} 
\usepackage{float}
\usepackage{epsfig,color,soul,rotating,overpic}
\usepackage{multicol}
\usepackage{multirow}
\usepackage[table,xcdraw]{xcolor}
\usepackage{wrapfig}
\usepackage[footnotesize,bf]{caption} 
\usepackage{lscape}
\usepackage{rotating}
\usepackage{pdflscape}
\usepackage{nomencl}
\newcolumntype{M}[1]{>{\centering\arraybackslash}m{#1}}

\usepackage{hyperref}
\hypersetup{
	colorlinks=true,
	linktoc=all,
	linkcolor=blue,
	citecolor=blue,
}

\definecolor{gray}{rgb}{0.5, 0.5, 0.5}
\definecolor{red}{rgb}{0.8500, 0.1250, 0.0480}
\definecolor{blue}{rgb}{0, 0.4470, 0.7410}

\title{Feedback control for transition suppression in direct numerical simulations of channel flow}

 \author{
  Yiyang Sun\thanks{Postdoctoral Associate, Department of Aerospace Engineering \& Mechanics, sun00249@umn.edu.}, \
   and Maziar S. Hemati\thanks{Assistant Professor, Department of Aerospace Engineering \& Mechanics, mhemati@umn.edu.} \\
      {\normalsize\itshape University of Minnesota, Minneapolis, MN 55455}
 } 

\usepackage{mdwlist}

\makecompactlist{itemize}{stditemize}

\usepackage{wrapfig}

\long\def\symbolfootnote[#1]#2{\begingroup%
\def\thefootnote{\fnsymbol{footnote}}\footnote[#1]{#2}\endgroup}

\begin{document}

\maketitle

\begin{abstract}

For channel flow at subcritical Reynolds numbers ($Re<5772$), a laminar-to-turbulent transition can emerge due to a large transient amplification in the kinetic energy of small perturbations, resulting in an increase in drag at the walls. The objectives of the present study are three-fold: (1)~to study the nonlinear effects on transient energy growth, (2)~to design a feedback control strategy to prevent this subcritical transition, and (3)~to examine the control mechanisms that enable transition suppression. We investigate transient energy growth of linear optimal disturbance in plane Poiseuille flow at a subcritical Reynolds number of $Re=3000$ using linear analysis and nonlinear simulation. We find that the amplification of the given initial perturbation is reduced when the nonlinear effect is substantial, with larger perturbations being less amplified in general. Moreover, we design linear quadratic optimal controllers to delay transition via wall-normal blowing and suction actuation at the channel walls. We demonstrate that these feedback controllers are capable of reducing transient energy growth in the linear setting. The performance of the same controllers is evaluated for nonlinear flows where a laminar-to-turbulent transition emerges without control. Nonlinear simulations reveal that the controllers can reduce transient energy growth and suppress transition.
Further, we identify and characterize the underlying physical mechanisms that enable feedback control to suppress and delay laminar-to-turbulent transition.

\end{abstract}

\section{Introduction}
\label{sec:intro}

Channel flow is ubiquitous in engineering applications, and the laminar-turbulent state of the flow is of great practical consequence. In general, since laminar flows experience smaller wall friction than turbulent flows, laminar-to-turbulent transition suppression has been a primary objective in many flow control studies of channel flow. Extensive studies have been performed to understand the characteristics of channel flows \cite{Stuart:JFM60, Watson:JFM60, Nishioka:JFM75, Henningson:JFM91, Elofsson:PF99}, among which hydrodynamic stability analysis is widely used to examine behaviors of small perturbations around a laminar equilibrium state of the flow. The linear stability result is often used to determine the critical condition for a laminar-to-turbulent transition to arise \cite{Thomas:PR53, Orszag:JFM71, Gustavsson:JFM91, Schmid01}. For plane Poiseuille flows, linear stability analysis identifies a critical Reynolds number of $Re=5772$, below which the flow state will remain laminar and infinitesimal perturbations will decay to zero asymptotically \cite{Orszag:JFM71}. However, a laminar-to-turbulent transition in plane Poiseuille flow has been observed in experiments and simulations at Reynolds numbers far smaller than the critical one \cite{Orszag:JFM80, Nishioka:JFM85, Chapman:JFM02}. This phenomenon is caused by a short-time transient amplification of small perturbations before their eventual decay. This so-called {\it transient energy growth} results from the non-normality of the linearized perturbation dynamics \cite{Schmid:ARF07, Hanifi:PF96}, which can contribute to the laminar-to-turbulent transition observed in plane Poiseuille flow at subcritical Reynolds numbers. 

To delay or suppress a transition in plane Poiseuille flow, modern feedback control theory has been investigated \cite{Joshi:JFM97, Bewley:JFM98, Hogberg:JF03, Ilak:AIAA08, Martinelli:PF11}.
One approach is to reduce transient energy growth by designing feedback control laws based on the linearized perturbation dynamics, as the large energy amplification plays an important role in the transition process.
  Kim and Bewley have summarized the essential ingredients of linear control theory for fluid mechanics in a review paper \cite{Kim:ARFM07}.
  Bewley and Liu applied linear optimal ($\mathcal H_2$) and robust ($\mathcal H_\infty$) control methods on both subcritical ans supercritical channel flows aiming at reducing disturbance response~\cite{Bewley:JFM98}.
  Martinelli et al.~designed a linear optimal control strategy to suppress the maximum transient energy growth experienced in subcritical channel flow~\cite{Martinelli:PF11}.
  Although the nonlinearity can be taken into consideration within the control synthesis~\cite{Jones:JFM15,Heins:Auto16}, linear control synthesis is more accessible and has been more commonly employed for flow control.

Among all the possible disturbances, the optimal disturbance leads to the maximum transient energy growth \cite{Schmid01}, which has the highest chance to induce a laminar-to-turbulent transition of the flow. The optimal disturbance is usually computed based on the linearized operator \cite{Whidborne:BIT11}, though nonlinear notions can also be determined \cite{Kerswell:ARFM18}. In the present work, we focus on the initial optimal disturbance computed based on the linear model.
The optimal disturbance is calculated for the uncontrolled and controlled systems, respectively. Control will alter the perturbation dynamics, and so the optimal disturbance for the controlled flow will differ from that of the uncontrolled flow, in general~\cite{Hemati:AIAAJ18}. This point has been emphasized in recent studies~\cite{Yao:AIAA18,Kalur:AIAA19,Yao:AIAA19} aiming to evaluate the overall performance of feedback controllers.

In this paper, we first perform linear and nonlinear simulations to study transient energy growth and transition in plane Poiseuille flows. We investigate nonlinear effects using direct numerical simulations.
In contrast to previous studies, we adopt a relatively large domain to resolve the flow and faithfully capture the full transition process.
  It is found that the nonlinearity serves to suppress the amplification of kinetic energy,
and that this effect becomes more pronounced for larger perturbations.
  For the oblique and streamwise-wave disturbances, streamwise vortices grow based on the development of coherent vortical structures initiated from the optimal disturbances. The breakdown of these streamwise vortices leads to a laminar-to-turbulent transition. For the spanwise-wave disturbance, high shear is induced between merged and large-scale streamwise vortices where secondary instabilities grow and break the coherent structure to trigger a laminar-to-turbulent transition. Although these transition mechanisms are not entirely new, we will focus on how feedback control favorably alters these mechanisms to impede the transition process.  

Linear quadratic controllers are designed to reduce the large transient energy growth experienced in the uncontrolled flows. Actuation is implemented in the form of wall blowing/suction within the nonlinear simulation, which has been shown capable of suppressing or delaying transition \cite{Martinelli:PF11}. Here, we investigate the control mechanism by which actuation favorably conditions the flow to prevent transition.  For the oblique and streamwise-wave disturbances, wall actuation modifies the distribution of high shear present in the flow, which prevents the large transient energy growth associated with the development of coherent structures. For the spanwise-wave disturbance, we will show that wall actuation induces small streamwise vortices near the channel wall, which hinders the growth of merged vortices and further reduces the high shear formed between adjacent streamwise vortical structures. 

The paper is organized as follows.
Numerical approaches are provided in Section \ref{sec:approach},
including details on the computational setup for direct numerical simulations and methods for feedback control synthesis.
In Section \ref{sec:result}, we discuss uncontrolled flows that exhibit transition for various optimal disturbances. The results for controlled flows are also provided in this section with detailed discussions of the control mechanisms. Finally, conclusions are summarized in Section \ref{sec:conclusion}.

\section{Numerical Approaches}
\label{sec:approach}

\subsection{Direct numerical simulation}
\label{sec:dns}

\begin{figure}[hbpt]
    \centering
    \includegraphics[width=0.65\textwidth]{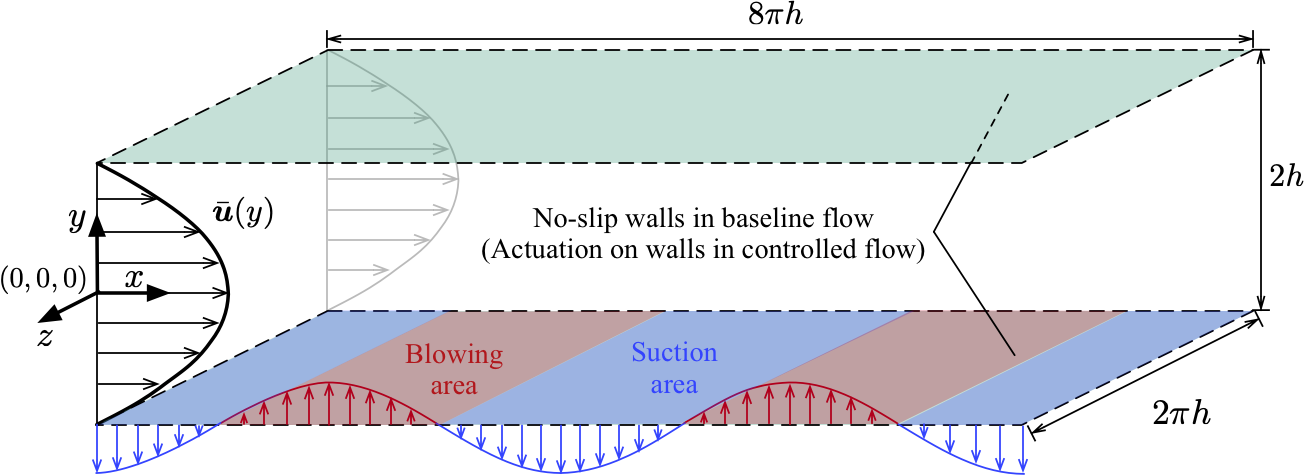}
    \caption{Schematic of plane Poiseuille flow with streamwise-wave disturbance (not to scale), both with and without wall actuation.}
    \label{fig:setup}
\end{figure}

Two- and three-dimensional direct numerical simulations of plane Poiseuille flow are performed using the spectral code {\it Channelflow} \cite{channelflow,GibsonHalcrowCvitanovicJFM08} to solve the incompressible Navier--Stokes equations. A second-order Runge--Kutta temporal scheme is used. Streamwise, wall-normal and spanwise directions are indicated by $x$, $y$, and $z$, respectively, and corresponding velocity components are $\boldsymbol u$, $\boldsymbol v$, and $\boldsymbol w$. Time is denoted by $t$.
As shown in Figure~\ref{fig:setup}, the flow between two infinite planes has a base velocity profile of $[\bar {\boldsymbol u}, \bar {\boldsymbol v}, \bar{\boldsymbol w}]=[1-(y/h)^2,0,0]$ at Reynolds number of $Re={\bar u_c h}/{\nu}=3000$,
where $h$ is the channel half-height, $\bar u_c$ denotes the center velocity of the base flow, and $\nu$ is the kinematic viscosity of the fluid.
The velocity is non-dimensionalized by $\bar u_c$. In the simulations, periodic boundary conditions are assumed in the $x$- and $z$-directions in which the flow variables (velocity and pressure) are represented by Fourier expansion. In the $y$-direction, flow variables are represented by Chebyshev polynomials, and no-slip boundary condition is specified at upper and lower walls for the uncontrolled flow. For the controlled flow, actuation in the form of blowing and suction in the $y$-direction is introduced on the entire upper and lower walls, which follows the Fourier expansion in the $x$- and $z$-directions as well.  The actuation is illustrated in Figure \ref{fig:setup} using an example of a case having streamwise-wavelike actuation. The amplitude of wall actuation is determined by the feedback law described later in Section \ref{sec:LQR}. 

We use a rectangular computational domain of size $8\pi h \times 2h \times 2\pi h$ in $x$-, $y$-, and $z$-directions, respectively. Chebyshev points of $N=101$ in the $y$-direction is prescribed to discretize the flow field for all the cases. Uniform grids of $128\times 64$ are used for the $x$- and $z$-directions, respectively. Although the cases with either streamwise-constant or spanwise-constant disturbances are two-dimensional flows, the laminar-to-turbulent transition is a three-dimensional flow phenomenon \cite{Reddy:JFM98} such that a three-dimensional mesh is necessary to resolve the transition process in the simulations. For both uncontrolled and controlled flows, grid resolution studies with doubled grids in each direction have been performed to ensure accuracy of the results.  Moreover, we have also studied the influences of computational domain size on the simulations results. In past studies, the domain size is usually set to fit one wavelength of the wavenumber of interest for each direction. However, as the transition includes large-scale deformation of original perturbation structures, here we select a relatively large domain to resolve the flow.
This enables a better understanding of the transition process through a more complete representation of the fluid dynamics.
  The evolution of the perturbation prior to transition is not affected by the domain size,
  but the post-transition process is dependent on the domain size.
  Once a sufficiently large domain size is determined, the simulations are converged and the observations among cases are consistent regardless of increasing domain size.

  For the simulations of both uncontrolled and controlled flows, the initial condition of the flow field consists of the base flow $[\bar {\boldsymbol u}, \bar {\boldsymbol v}, \bar {\boldsymbol w}]$ and a small initial perturbation $[\boldsymbol{u}'_0,\boldsymbol{v}'_0,\boldsymbol{w}'_0]$, where subscript $(\cdot)_0$ denotes initial value. The perturbation is the optimal disturbance associated with a wavenumber pair of $(\alpha$,$\beta)$, and its kinetic energy density is denoted by $E_0$. The optimal disturbance can lead to a maximum transient energy growth and it is pre-calculated using the algorithm proposed by Whidborne and Amar~\cite{Whidborne:BIT11}. Since the dynamical systems of the uncontrolled and controlled flows are different, the optimal disturbance is calculated for each system independently to guarantee that we examine the largest transient energy growth in each case.  Moreover, a random perturbation with kinetic energy density of $1\%$ of $E_0$ is added to expedite the emergence of a laminar-to-turbulent transition in the flows. 
The addition of a random perturbation has a negligible influence on the shape of the optimal disturbance, and is a common practice in transition studies~\cite{Reddy:JFM98}.

\subsection{Linearized Navier--Stokes equations and feedback control design}
\label{sec:LQR}

To use modern control theory to design feedback controllers, a state-space representation of the fluid flow problem is required. For plane Poiseuille flow, we decompose the flow state $\tilde {\boldsymbol q}$ into a base state $\bar {\boldsymbol{q}}$ and a small perturbation $\boldsymbol{q}'$, where $\boldsymbol{q}=[\boldsymbol{u}, \boldsymbol{v},\boldsymbol{w},\boldsymbol{p}]^T$ ($\boldsymbol{p}$ is pressure), and the base state velocity profile is of $[\bar {\boldsymbol u}, \bar{\boldsymbol v}, \bar {\boldsymbol w}]=[1-(y/h)^2,0,0]$. The kinetic energy density of a perturbation is defined as 
\begin{equation}
E=\frac{1}{2V}\int_{vol=V} (\boldsymbol{u}'^2+\boldsymbol{v}'^2+\boldsymbol{w}'^2) dvol,
\end{equation}
where $V$ is the volume of the computational domain.

By substituting the expression of $\tilde {\boldsymbol q}=\bar {\boldsymbol q}+\boldsymbol{q}'$ into the Navier--Stokes equations, and assuming that the perturbation is much smaller than the base state in magnitude ($|\boldsymbol{q}'|\ll |\bar{\boldsymbol{q}}|$), we linearize the equations by retaining linear terms and neglecting higher-order nonlinear terms. Next, the real-valued perturbation is expressed using a Fourier expansion   
\begin{equation} 
\boldsymbol{q}'(x,y,z,t) = \hat {\boldsymbol q}(y,t)e^{i(\alpha x +\beta z)} + \text{complex conjugate},
\label{modal}
\end{equation}
where $\hat {\boldsymbol q}(y,t)$ is the amplitude function of the perturbation associated with streamwise wavenumber $\alpha$ and spanwise wavenumber $\beta$. By substituting Eq.~(\ref{modal}) into the linearized Navier--Stokes equations, we can reformulate the governing equations into a state-space form 
\begin{equation}
\frac{\partial X}{\partial t} = {A}(\bar {\boldsymbol q}; \alpha, \beta)  {X},
\label{OL}
\end{equation} 
where ${X} = [\hat {\boldsymbol v}, \hat {\boldsymbol \eta}]^T$, $\hat {\boldsymbol v}$ and $\hat {\boldsymbol \eta}$ denote wall-normal velocity perturbation and wall-normal vorticity perturbation, respectively, and superscript $(\cdot)^T$ represents transpose. The dynamics matrix $A$ is derived from linearized Navier--Stokes equations, which follows the form proposed by Schmid \& Henningson \cite{Schmid01}. By choosing a real-valued wavenumber pair ($\alpha$, $\beta$), we can analyze perturbations associated with various structures. In the present work, we examine optimal disturbances for three different wavenumber pairs --- $(\alpha,\beta)=(1,0)$, (0,2) and (1,1). The features of each optimal disturbance will be discussed in Section \ref{sec:result}.

In the control design, we introduce actuation in the form of blowing and suction on the upper and lower channel walls. To synthesize this actuation into the dynamical system, we add a control input term $BU$ to the uncontrolled system (Eq.~(\ref{OL})) to form the system
\begin{equation}
\frac{\partial X_c}{\partial t}  =  A_cX_c + BU,
\label{ss_CL}
\end{equation}
where the control inputs in ${U}=\frac{\partial}{\partial t}[\hat v_{+h}, \hat v_{-h}]^T$ are the rate of change of wall-normal velocity on the upper and lower walls, and ${B}$ is the input matrix that maps the influence of control inputs on the system. Due to the no-slip boundary condition ($\hat v_{\pm h}=0$) applied on the walls of the uncontrolled flow, $\hat v_{\pm h}$ are excluded while forming the flow state $X$ in Eq.~(\ref{OL}). As $\hat v_{\pm h}$ are nonzero in the controlled flow, we will append these two variables to the flow state $X$ to form a new state $X_c=[X, \hat v_{+h}, \hat v_{-h}]^T$. Analogously, the dynamics matrix $A$ is modified and denoted by $A_c$ to account for the new state.  More information about the system modeling can be found in the work by McKernan et al.~\cite{McKernan:IMIC06}. 

Based on the state-space model in Eq. (\ref{ss_CL}), linear-quadratic regulator (LQR) is designed for use in feedback control. The control objective is to minimize the cost function
\begin{equation}
J=\int ^\infty _0 ( {X_c}^T  {Q}  {X_c}+ {U}^T  {R}  {U})dt
\end{equation}
subject to the linear dynamics (Eq.~(\ref{ss_CL})), where $Q$ and $R$ are weight matrices. The term $X_c^TQX_c$ represents kinetic energy density $E$.
The input weighting matrix $R$ serves to penalize the control effort and is chosen here to be a diagonal matrix with each element equal to $10^{-6}$.
The feedback control law is given by $U=-KX_c$, where matrix $K$ is the state feedback control gain. Given the dynamics matrix $A_c$, input matrix $B$, weight matrices $Q$ and $R$, the control gain matrix $K$ can be calculated by solving an algebraic Riccati equation \cite{Brogan91}.
The LQR controller is implemented in the direct numerical simulations, as described earlier. Results are presented in the next section.

\section{Results and Discussions}
\label{sec:result}

In this section, transient energy growth of optimal disturbances associated with three wavenumber pairs of $(\alpha,\beta)=(1,0)$, $(0,2)$ and $(1,1)$ are investigated using direct numerical simulations for the subcritical plane Poiseuille flow at $Re=3000$. The required conditions for a laminar-to-turbulent transition to appear in a fully nonlinear flow are examined for each case, and the flow features that are responsible for the transition are examined as well. Moreover, we present LQR feedback controllers that can successfully suppress or delay the transition, then examine the underlying physics related to the control mechanisms.   

\subsection{Transient energy growth of optimal disturbance}

\begin{figure}[hbpt]
    \centering
    \includegraphics[width=1.0\textwidth]{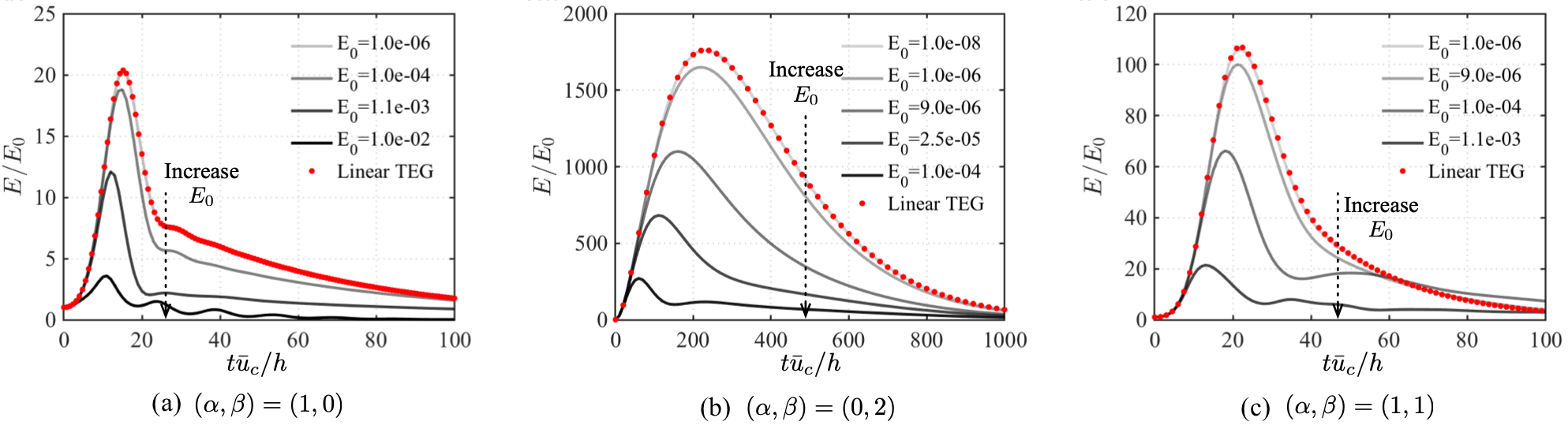}
    \caption{Transient energy growth (TEG) of optimal disturbance with a range of initial kinetic energy density $E_0$ from linear (red dots) and nonlinear (solid lines) simulations for uncontrolled flows.}
    \label{fig:teg_base}
\end{figure}

We start the discussion from the nonlinear analysis of transient energy growth resulting from a pure optimal disturbance---i.e.,~without a random perturbation added. Using the optimal disturbance as the initial condition, the Navier--Stokes equations were integrated to obtain the response of kinetic energy density, as shown in Figure~\ref{fig:teg_base}.
All of the optimal disturbances considered in the present work experience large transient energy amplifications with orders of magnitude from $O(10^{1})$ to $O(10^{3})$; the disturbance of streamwise vortices ($(\alpha, \beta)=(0,2)$ in Figure~\ref{fig:teg_base}(b) ) reaches the largest maximum transient energy growth.
In the linear simulation---denoted by red dots in each subplot---the normalized energy response $E/E_0$ does not depend on the magnitude of the initial perturbation.

In the nonlinear simulations, we integrate the Navier--Stokes equations of full state velocity $\tilde {\boldsymbol q}$ and report the kinetic energy density of perturbations after subtracting the base flow $\bar {\boldsymbol q}$. As shown in Figure~\ref{fig:teg_base}, when the perturbation is small enough such as $E_0\lessapprox 10^{-6}$ for cases with $(\alpha,\beta)=(1,0)$ and $(1,1)$ and $E_0\lessapprox 10^{-8}$ for case with $(\alpha,\beta)=(0,2)$, nonlinear effects are negligible and the energy response is identical to that of the linear result. However, when the amplitude of initial disturbance increases, the amplification of kinetic energy density $E/E_0$ decreases, which indicates that the nonlinearity suppresses energy amplification. In other words, the nonlinearity competes with the linear non-modal growth and inhibits the transient energy amplification for sufficiently large $E_0$.

Although we have seen transient energy growth in both linear and nonlinear simulations, there is no laminar-to-turbulent transition observed, suggesting that the growth of optimal disturbance alone cannot trigger the transition. As discussed in the study by Reddy et al.~\cite{Reddy:JFM98}, the addition of a random perturbation is required to observe transition in simulations. Hence, we added a random perturbation with a contribution of $1\%$ of $E_0$ to the optimal disturbance to initiate laminar-to-turbulent transition.
Before the laminar-to-turbulent transition, the energy response due to this modified disturbance is nearly identical to the energy response due to a purely optimal disturbance.
Since the transition process of each wavenumber pair follows different routes, we will discuss all the three paths to transition below respectively.

\subsubsection{Oblique disturbance $(\alpha,\beta)=(1,1)$}
\label{sec:base11}

\begin{figure}[hbpt]
    \centering
    \includegraphics[width=1.0\textwidth]{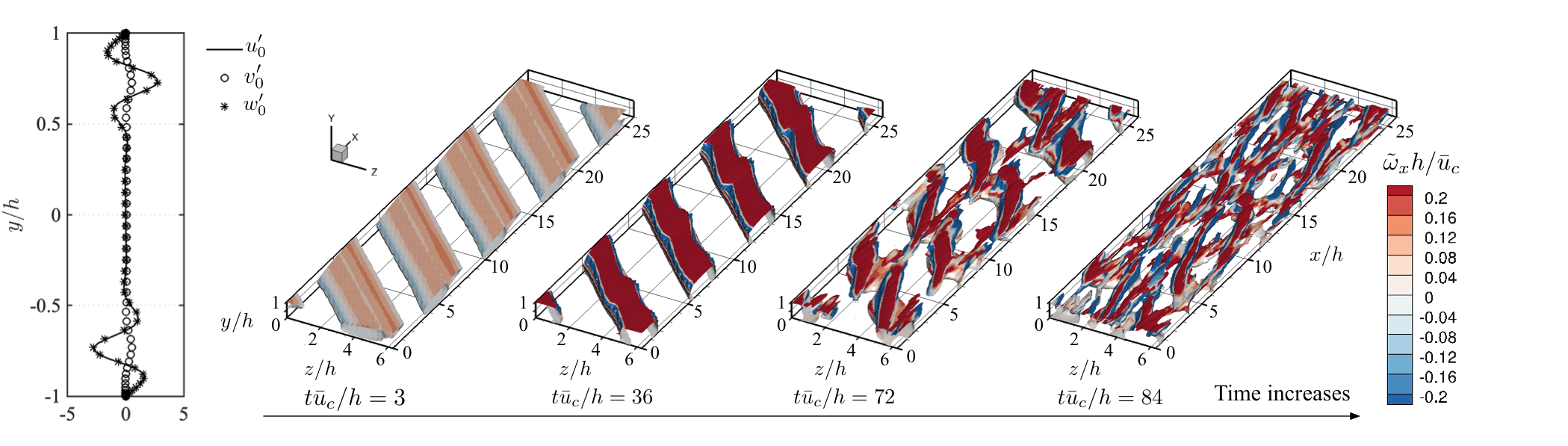}
    \caption{Time evolution of flow field to illustrate the laminar-to-turbulent transition process for the case $(\alpha,\beta)=(1,1)$. The flow is initialized by optimal disturbance with kinetic energy density of $E_0=0.5\times 10^{-4}$. Iso-surfaces of $\mathcal Q$-criterion \cite{Hunt:88} ($\mathcal Q(h/\bar u_c)^2=0.001$) colored by streamwise vorticity $\tilde \omega_x h/\bar u_c$ are visualized. Only the upper half domain ($y\ge 0$) is displayed for clarity.}
    \label{fig:dns_base_A1B1_flowfield}
\end{figure}

\begin{figure}[hbpt]
    \centering
    \includegraphics[width=0.9\textwidth]{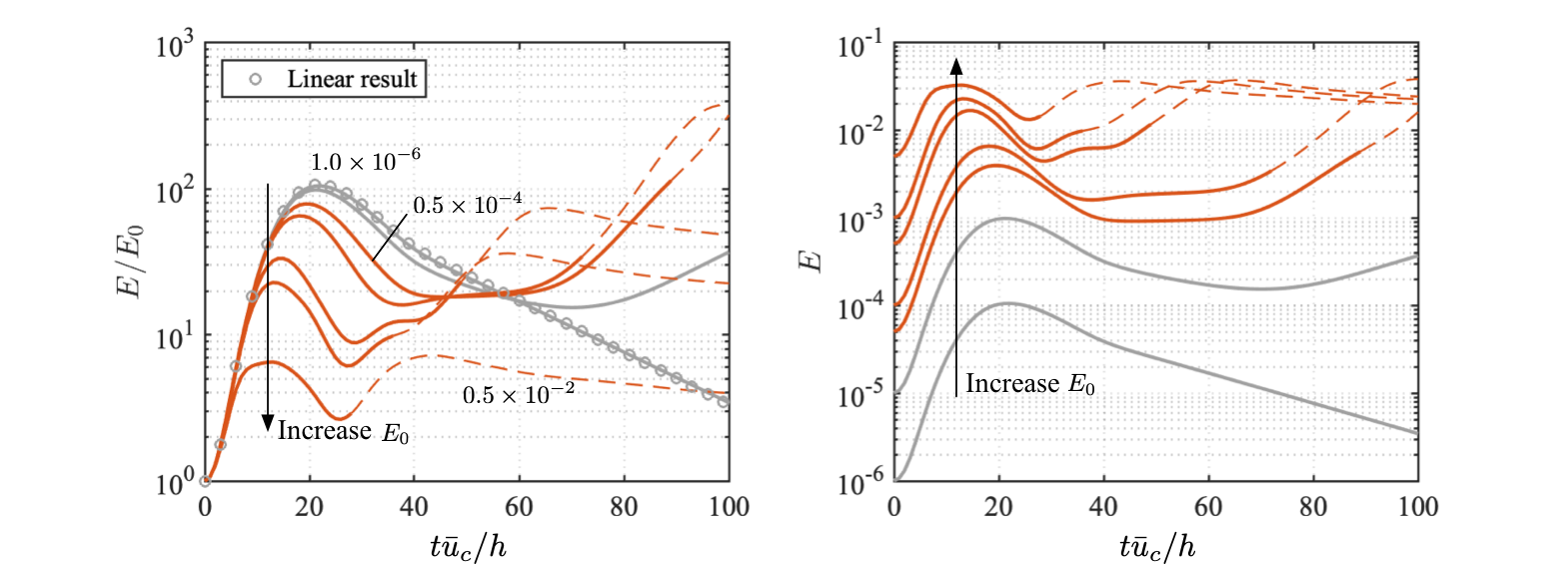}
    \caption{Time evolution of kinetic energy density $E$ of case $(\alpha,\beta)=(1,1)$ initialized by different amplitudes of optimal disturbance $E_0$. Red lines represent transition cases, and dashed line indicates that the flow is in turbulent state.}
    \label{fig:dns_base_A1B1_E}
\end{figure}

Flow fields of a laminar-to-turbulent transition case with $(\alpha,\beta)=(1,1)$ and $E_0=0.5\times 10^{-4}$ are shown in Figure~\ref{fig:dns_base_A1B1_flowfield}, and the iso-surface of $\mathcal Q$-criterion \cite{Hunt:88} is calculated using full flow state $\tilde {\boldsymbol q}$. The initial optimal disturbance is in an oblique-wave with large amplitude near the channel walls, as shown in the subplot in Figure~\ref{fig:dns_base_A1B1_flowfield}. As the disturbance evolves in time, large-scale oblique structures form with size comparable to the channel half-height. After the transient energy reaches to the maximum value around $t\bar u_c/h=20$ (see Figure \ref{fig:dns_base_A1B1_E}), large amplitude of streamwise vorticity appears in the vicinity of the coherent structures at the channel walls at $t\bar u_c/h=36$. Meanwhile, slight spatial variations of the coherent structures appear along the oblique direction. Later at time $t\bar u_c/h=72$, streamwise vortices are induced from the oblique coherent structures and grow into streamwise streaks, although the overall transient energy density has started to decay, as shown in Figure~\ref{fig:dns_base_A1B1_E}. After $t\bar u_c/h=84$, these streamwise streaks break into small-scale structures, and a laminar-to-turbulent transition emerges afterwards.

The time evolution of kinetic energy density $E$ for a range of optimal disturbances is presented in Figure \ref{fig:dns_base_A1B1_E}. For optimal disturbance with $E_0< 0.5\times 10^{-4}$ (denoted by grey lines), the initial disturbance is the most amplified, yet the induced streamwise vortices do not grow significantly. The disturbance eventually decays to zero without triggering a laminar-to-turbulent transition. We also note that the nonlinear flow with small disturbance ($E_0=1.0\times 10^{-6}$) is well predicted by the linear analysis as seen by the overlap between nonlinear simulation and linear results in Figure \ref{fig:dns_base_A1B1_E} (left).  By examining the absolute value of the perturbation kinetic energy (Figure \ref{fig:dns_base_A1B1_E} (right)), the transition cases have relatively high values in $E$, which suggests that a combined effect of initial optimal disturbance and its corresponding amplification determines whether or not a transition occurs. The larger amplitude disturbance gives a higher chance of triggering transition to turbulent state, which also indicates that a reduction in transient energy growth has potential to suppress or delay the transition observed in the uncontrolled flow. This motivates using control to reduce transient energy growth.

\begin{figure}[hbpt]
    \centering
    \includegraphics[width=0.4\textwidth]{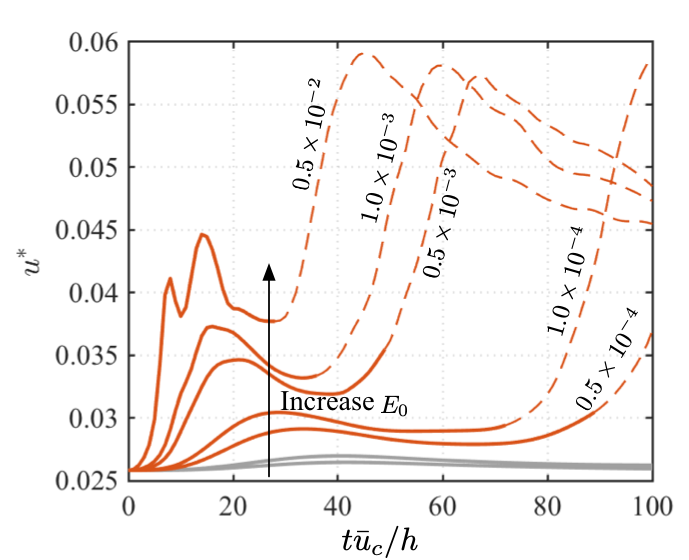}
    \caption{Time evolution of friction velocity $u^*$ of case $(\alpha,\beta)=(1,1)$ initialized by different amplitudes of optimal disturbance $E_0$. Red lines represent transition cases, and grey lines denote that the flow remains laminar. Dashed line indicates that the flow is in turbulent state.}
    \label{fig:dns_base_A1B1}
\end{figure}

Friction velocity $u^*$ is defined using wall shear stress $\tau_w$ and density of the flow
\begin{equation}
u^*=\sqrt{\frac{\tau_w}{\rho}},    
\end{equation}
where $\tau_w=\mu (\partial u/\partial y)$, $\mu$ is dynamic viscosity, and density $\rho$ is set to one for incompressible flow. As a sharp increase in friction velocity $u^*$ indicates an emergence of laminar-to-turbulent transition, a summary of transition cases are shown in Figure \ref{fig:dns_base_A1B1} illustrating the features of friction velocity $u^*$, in which the flow in turbulent state has been denoted by dashed lines. Based on the feature of friction velocity, the larger the initial disturbance, the earlier laminar-to-turbulent transition appears. Moreover, a large initial disturbance leads to large shear stress in the wall-normal direction near the channel walls. 

\subsubsection{Streamwise-wave disturbance $(\alpha,\beta)=(1,0)$}
\label{sec:base10}

\begin{figure}[hbpt]
    \centering
    \includegraphics[width=1.0\textwidth]{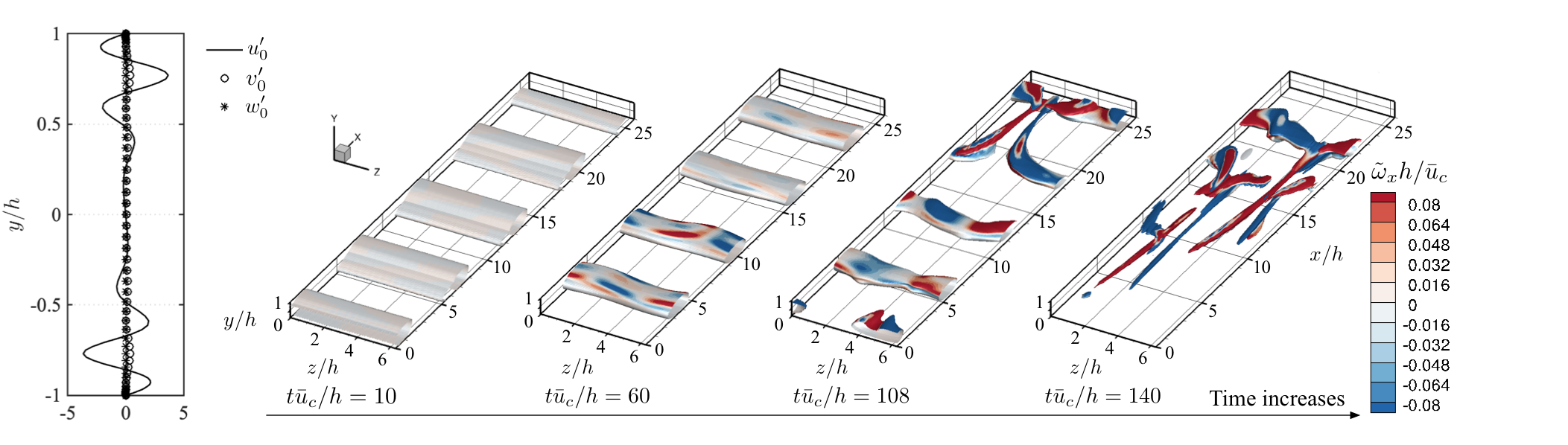}
    \caption{Time evolution of flow field to illustrate the laminar-to-turbulent transition process for the case $(\alpha,\beta)=(1,0)$. The flow is initialized by optimal disturbance with kinetic energy density of $E_0=1.0\times 10^{-4}$. Iso-surfaces of $\mathcal Q$-criterion ($\mathcal Q(h/\bar u_c)^2=0.005$) colored by streamwise vorticity $\tilde \omega_x h/\bar u_c$ are visualized. Only the upper half domain ($y\ge 0$) is displayed for clarity.}
    \label{fig:dns_base_A1B0_flowfield}
\end{figure}

The optimal disturbance with wavenumber pair of $(\alpha,\beta)=(1,0)$ is the well-known Tollmien-Schlichting wave whose structures are uniform in the spanwise direction. The evolution of the optimal disturbance is shown in Figure \ref{fig:dns_base_A1B0_flowfield} with $E_0=1.0\times 10^{-4}$. The feature of the optimal disturbance is similar to the case with oblique disturbance: a large disturbance is observed in streamwise velocity and resides near the channel walls. As the optimal disturbance grows into large-scale spanwise coherent structures, streamwise vorticity increases near channel walls, especially where the coherent structures reside, as seen in Figure \ref{fig:dns_base_A1B0_flowfield} at $t\bar u_c/h=60$. When the streamwise vorticity is large enough, a $\Lambda$-shaped structure emerges, splits the spanwise coherent structure and connects to the adjacent spanwise coherent structure in front, as seen at $t\bar u_c/h=108$, then the break of these vortices leads to a laminar-to-turbulent transition of the flow. 

\begin{figure}[hbpt]
    \centering
    \includegraphics[width=0.45\textwidth]{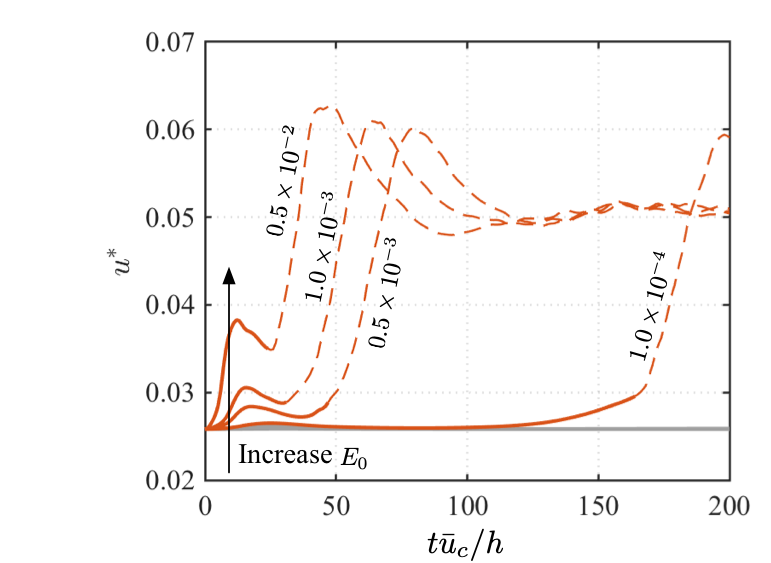}
    \caption{Time evolution of friction velocity $u^*$ of case $(\alpha,\beta)=(1,0)$ initialized by different amplitudes of optimal disturbance $E_0$. Red lines represent transition cases, and grey lines denote that the flow remains laminar. Dashed line indicates that the flow is in turbulent state.}
    \label{fig:dns_base_A1B0}
\end{figure}

\begin{figure}[hbpt]
    \centering
    \includegraphics[width=0.9\textwidth]{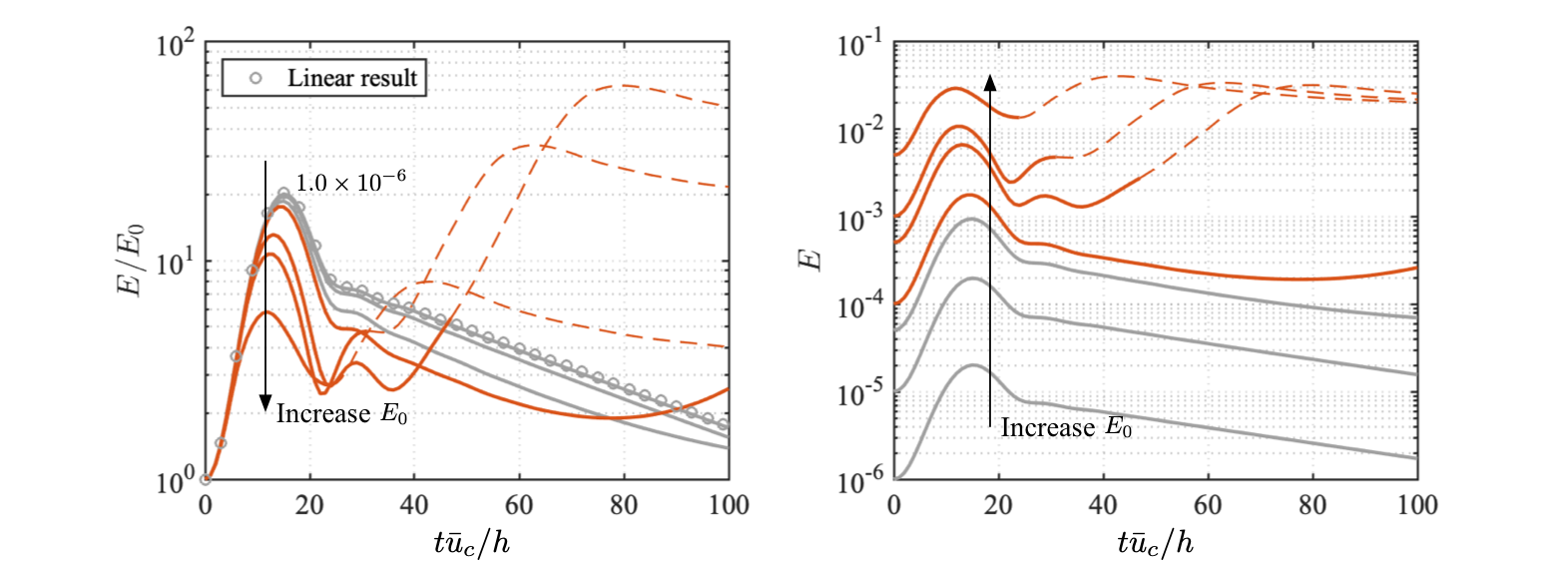}
    \caption{Time evolution of kinetic energy density $E$ of case $(\alpha,\beta)=(1,0)$ initialized by different amplitudes of optimal disturbance $E_0$. Red lines represent transition cases, and grey lines denote that the flow remains laminar. Dashed line indicates that the flow is in turbulent state.}
    \label{fig:dns_base_A1B0_E}
\end{figure}

The transition process in this case is very similar to the oblique disturbance discussed above in Section \ref{sec:base11}. The transition is directly caused by a break of streamwise streaks that grow from the amplified optimal disturbance. It suggests that the transient energy growth of the optimal disturbance alone does not trigger the transition; rather, it is the combined effect of the growing optimal disturbance and induced secondary instabilities that leads to a laminar-to-turbulent transition. The friction velocity and evolution of transient energy density of disturbance with $(\alpha,\beta)=(1,0)$ are shown in Figures \ref{fig:dns_base_A1B0} and \ref{fig:dns_base_A1B0_E}, respectively. The features observed from the case $(\alpha,\beta)=(1,1)$ also apply on this streamwise-wave disturbance; the flow tends to transition into a turbulent state when the kinetic energy density $E$ of a disturbance becomes large; this is influenced by both the initial amplitude of the disturbance and the corresponding amplification rate.

\subsubsection{Spanwise-wave disturbance $(\alpha,\beta)=(0,2)$}
\label{sec:base02}

\begin{figure}[hbpt]
    \centering
    \includegraphics[width=1.0\textwidth]{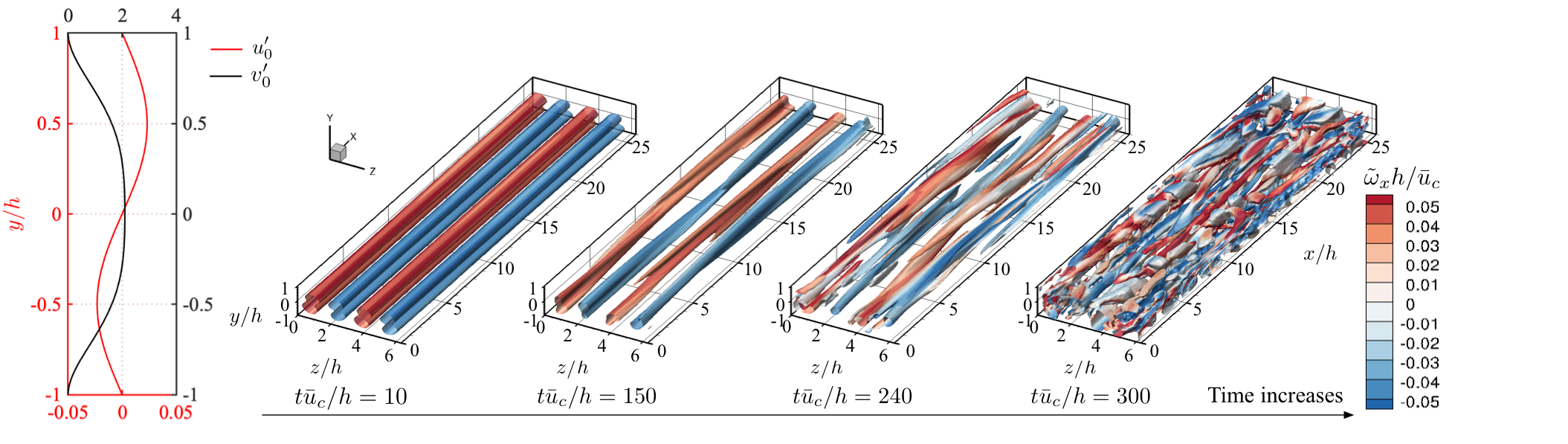}
    \caption{Time evolution of flow field to illustrate the laminar-to-turbulent transition process for the case $(\alpha,\beta)=(0,2)$. The flow is initialized by optimal disturbance with kinetic energy density of $E_0=1.0\times 10^{-4}$. Iso-surfaces of $\mathcal Q$-criterion ($\mathcal Q(h/\bar u_c)^2=0.0004$) colored by streamwise vorticity $\tilde \omega_x h/\bar u_c$ are visualized.}
    \label{fig:dns_base_A0B2_flowfield}
\end{figure}

\begin{figure}[hbpt]
    \centering
    \includegraphics[width=0.9\textwidth]{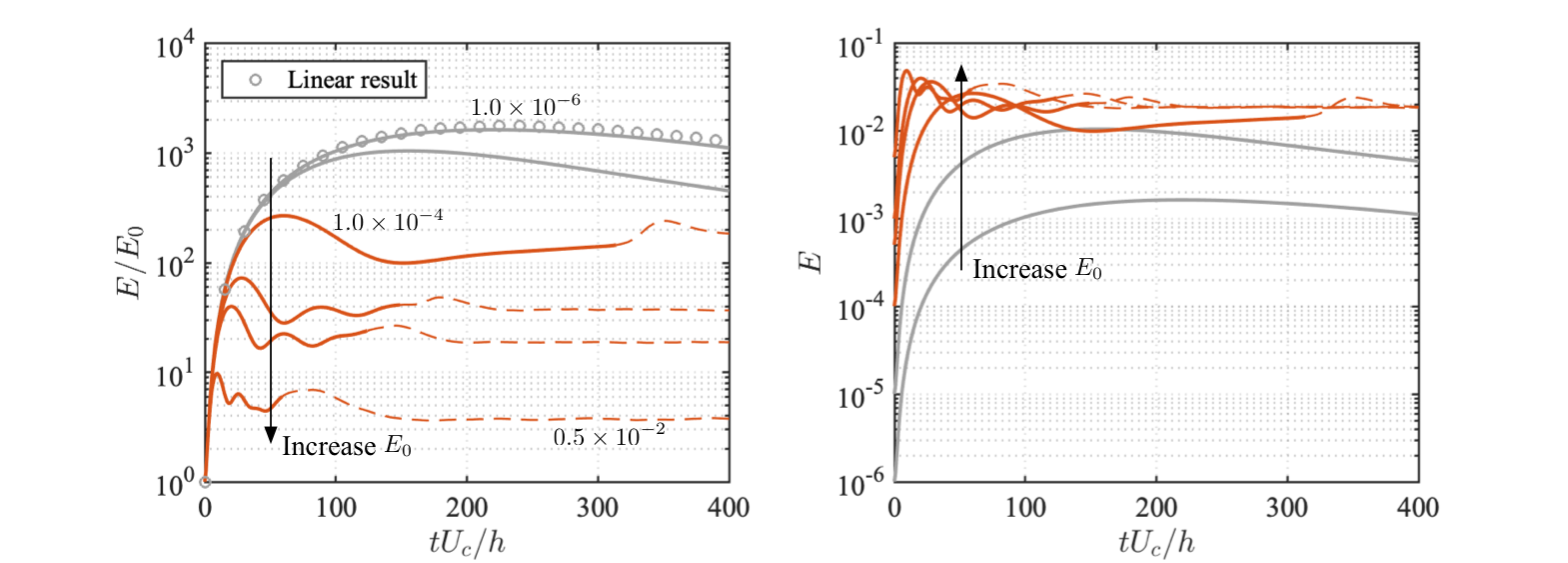}
    \caption{Time evolution of kinetic energy density $E$ of case $(\alpha,\beta)=(0,2)$ initialized by different amplitudes of optimal disturbance $E_0$. Red lines represent transition cases, and grey lines denote that the flow remains laminar. Dashed line indicates that the flow is in turbulent state.}
    \label{fig:dns_base_A0B2_E}
\end{figure}

\begin{figure}[hbpt]
    \centering
    \includegraphics[width=0.45\textwidth]{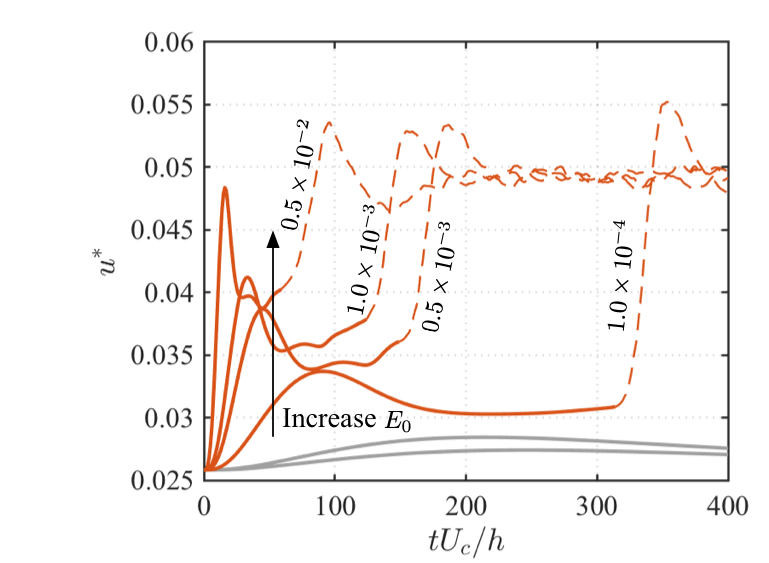}
    \caption{Time evolution of friction velocity $u^*$ of case $(\alpha,\beta)=(0,2)$ initialized by different amplitudes of optimal disturbance $E_0$. Red lines represent transition cases, and grey lines denote that the flow remains laminar. Dashed line indicates that the flow is in turbulent state.}
    \label{fig:dns_base_A0B2}
\end{figure}

The optimal disturbance associated with wavenumber pair of $(\alpha,\beta)=(0,2)$ is in the form of streamwise vortices. Shown in Figure \ref{fig:dns_base_A0B2_flowfield} are the flow fields with initial kinetic energy density of $E_0=1.0\times 10^{-4}$. Relating the flow feature to normalized kinetic energy density of perturbation $E/E_0$ shown in Figure \ref{fig:dns_base_A0B2_E} (left), when the optimal disturbance grows initially, two co-rotating vertically aligned vortices merge into a large streamwise vortical structure. The structure remains uniform in the streamwise direction, and the corresponding transient energy reaches the maximum value around $t\bar u_c/h=60$. After the kinetic energy starts to decrease, the flow gradually experiences spatial variation as seen at $t\bar u_c/h=150$. This is associated with axial rotational motion around the streamwise vortical structures with comparable scale to the length of the streamwise domain. Next, the kinetic energy density of the flow increases again after $t\bar u_c/h=150$, and smaller structures appear in the flow (as seen at $t\bar u_c/h=240$). After a short duration of rotational motion in the flow, a laminar-to-turbulent transition occurs around $t\bar u_c/h=300$. Based on the features of transient energy growth and corresponding flow fields, we note that the transition is a slow process covering several stages, which cannot be determined simply based on the features of energy growth time histories. The features of transition cases revealed by friction velocity are similar to the cases of oblique disturbances and streamwise-wave disturbances discussed above, and the friction velocity of case $(\alpha,\beta)=(0,2)$ is shown in Figure \ref{fig:dns_base_A0B2}. 

\subsection{Suppression of laminar-to-turbulent transition}

We synthesize a full-state feedback LQR controller. The transient energy growth is significantly reduced with the corresponding wall actuation, as shown in Figure \ref{fig:teg_ctr}. In the nonlinear simulations, we update the control input $U$ at each time step using the flow state $X_c$ and the gain matrix $K$ calculated from the linear analysis. Accordingly, the boundary condition of wall-normal velocity $v'_{\pm h}$ is modified based on the control input $U$. To evaluate the overall performance of the controlled flow in terms of reducing large transient energy growth, the optimal disturbance was re-calculated for the closed-loop controlled system to evaluate the maximum transient energy growth.

\subsubsection{Reduction in transient energy growth}

\begin{figure}[hbpt]
    \centering
    \includegraphics[width=1.0\textwidth]{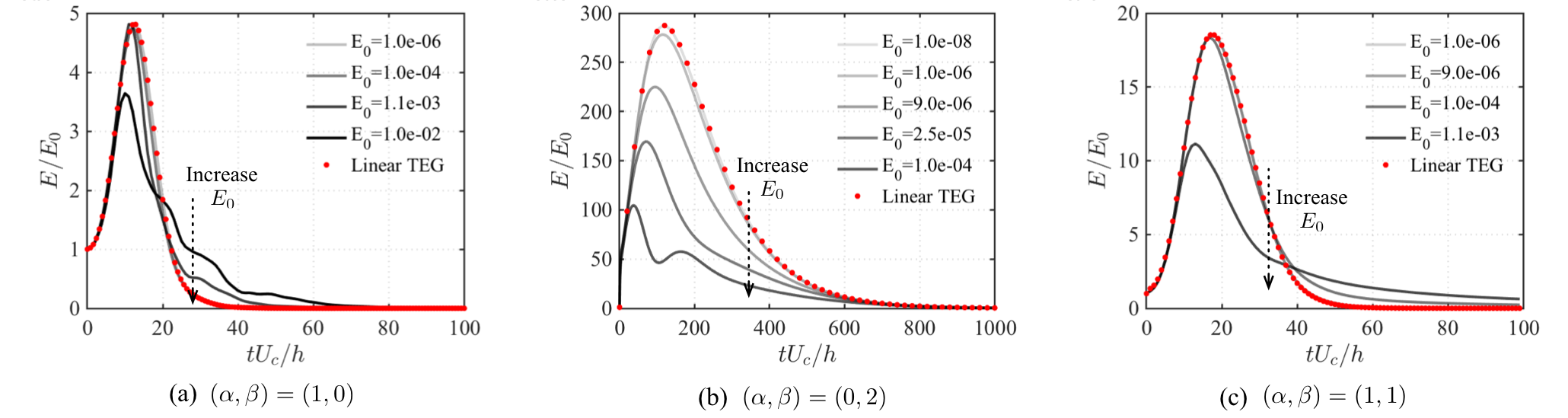}
    \caption{Transient energy growth (TEG) of optimal disturbance with a range of initial kinetic energy density $E_0$ from linear (red dots) and nonlinear (solid lines) simulations for controlled flows.}
    \label{fig:teg_ctr}
\end{figure}

Transient energy growth in each controlled case is observed to exhibit similar features as in the uncontrolled cases; namely that an increase in initial kinetic energy makes the amplification of perturbation energy decrease as shown in Figure \ref{fig:teg_ctr}. Using case $(\alpha,\beta)=(1,0)$ as an example, since the amplification of the optimal disturbance is effectively reduced with the feedback controller, the energy responses with $E_0\lessapprox 1.1\times 10^{-3}$ almost overlap the linear result, denoted by red dots.  This suggests that, as the growth of the disturbance is suppressed to a small magnitude by introducing control, the importance of the nonlinearity is weakened. Therefore, the results from nonlinear simulations match the result from linear analysis as the small disturbance assumption is satisfied.   

\begin{figure}[hbpt]
    \centering
    \includegraphics[width=0.65\textwidth]{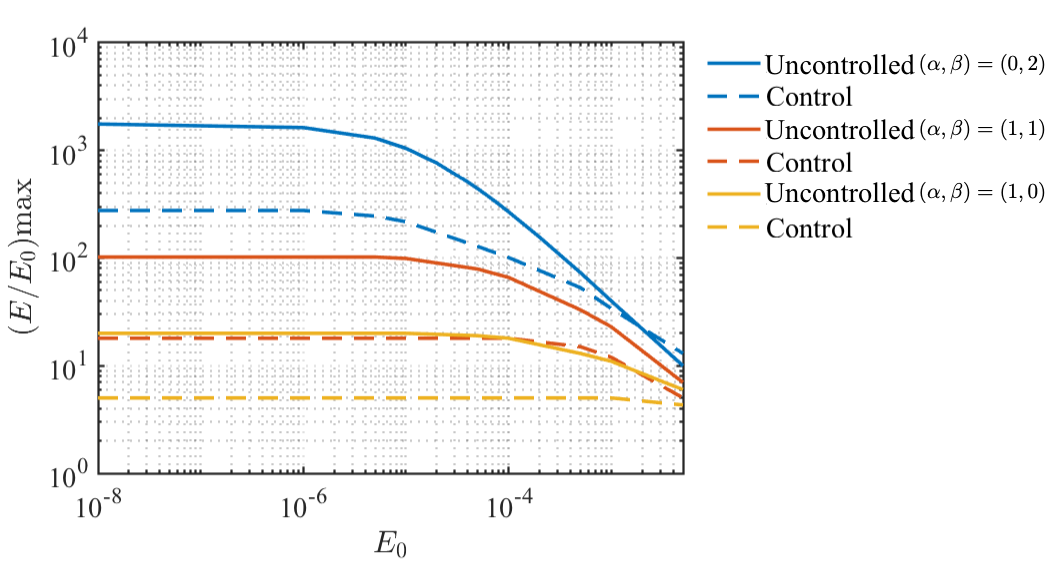}
    \caption{A comparison of maximum amplified transient energy density $(E/E_0)_\text{max}$ of a range of amplitude of initial optimal disturbance between uncontrolled (solid lines) and controlled (dashed lines) cases.}
    \label{fig:teg_ctr_percent}
\end{figure}

For each wavenumber pair, a comparison of the maximum of $E/E_0$ initiated from a range of optimal disturbance amplitudes is shown in Figure \ref{fig:teg_ctr_percent}. For small disturbance amplitude, the percent reduction of maximum transient energy remains nearly identical, indicating the nonlinearity of small disturbance is negligible as assumed in the linear analysis. Moreover, each wavenumber scenario reaches the largest reduction of $\approx 80\%$ in $(E/E_0)_\text{max}$ for small initial amplitude of disturbance. For the case with $(\alpha,\beta)=(0,2)$, this large amount of reduction in $E/E_0$ only applies to the cases with $E_0\lessapprox O(10^{-5})$. As the initial disturbance grows in amplitude, the control performance in terms of suppressing the energy amplification degrades, such that the $(E/E_0)_\text{max}$ is nearly unchanged for cases with $E_0 > O(10^{-3})$.  Similar findings are also observed in the other two disturbance scenarios, but the reduction in $(E/E_0)_\text{max}$ is maintained for $E_0\lessapprox O(10^{-4})$ and $E_0\lessapprox O(10^{-3})$ for case $(\alpha,\beta)=(1,1)$ and $(\alpha,\beta)=(1,0)$, respectively. 

\subsubsection{Controlled flow with oblique or streamwise-wave disturbances}

By reducing the transient energy growth using feedback control, the laminar-to-turbulent transition experienced in the flow with oblique or streamwise-wave disturbances has been successfully suppressed. The threshold becomes higher in terms of triggering the transition process. Here, we examine how the wall actuation modifies the flow characteristics to prevent the transition in the uncontrolled flow. Because the control mechanism is similar in the cases with oblique and streamwise-wave disturbances, i.e., $(\alpha,\beta)=(1,1)$ and $(1,0)$, respectively, we will use the flow with oblique disturbance as an example to illustrate the control mechanism.   

\begin{figure}[hbpt]
    \centering
    \includegraphics[width=0.4\textwidth]{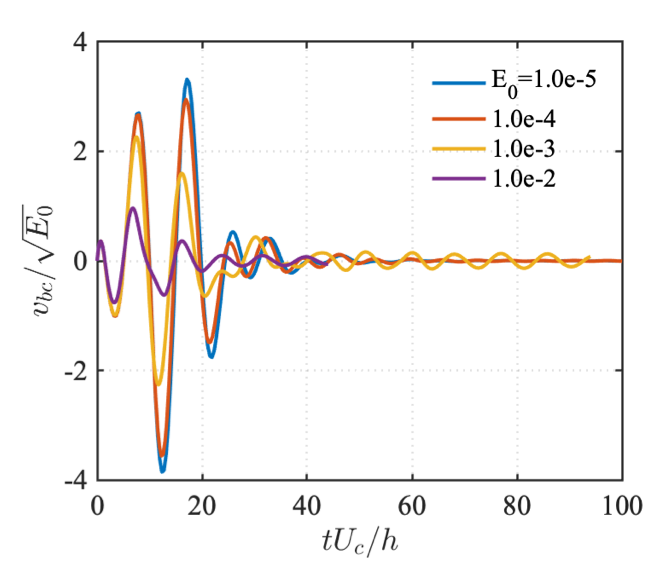}
    \caption{Time history of normalized wall-normal velocity $v_\text{bc}/\sqrt{E_0}$ at lower wall of case $(\alpha,\beta)=(1,1)$ when actuation is turned on. }
    \label{fig:A1B1_vbc}
\end{figure}

As the wall actuation is in the form of Fourier waves in space, here we show the amplitude of the wall-normal velocity at $[x,y,z]/h=[0,-1,0]$ to illustrate the actuation time-history in Figure \ref{fig:A1B1_vbc}. The wall-normal velocity is normalized by the corresponding $\sqrt{E_0}$. For case with $E_0=1.0\times 10^{-4}$, the transition is successfully suppressed by introducing the wall actuation. The wall-normal velocity is similar to the actuation history of smaller disturbance case ($E_0=1.0\times 10^{-5}$). Large blowing and suction are applied initially during time $0< t\bar u_c/h <30$ when large transient energy growth is present in the flow. Once the growth has been suppressed, the wall actuation gradually decays to zero. 
For the cases with larger perturbations ($E_0=1.0\times 10^{-3}$ and $1.0\times 10^{-2}$), the linear feedback control strategy fails to suppress the transition, and the flow transitions to a turbulent state. The normalized wall-normal velocity deviates from the successfully controlled cases and oscillates longer in time.

\begin{figure}[hbpt]
    \centering
    \includegraphics[width=1.0\textwidth]{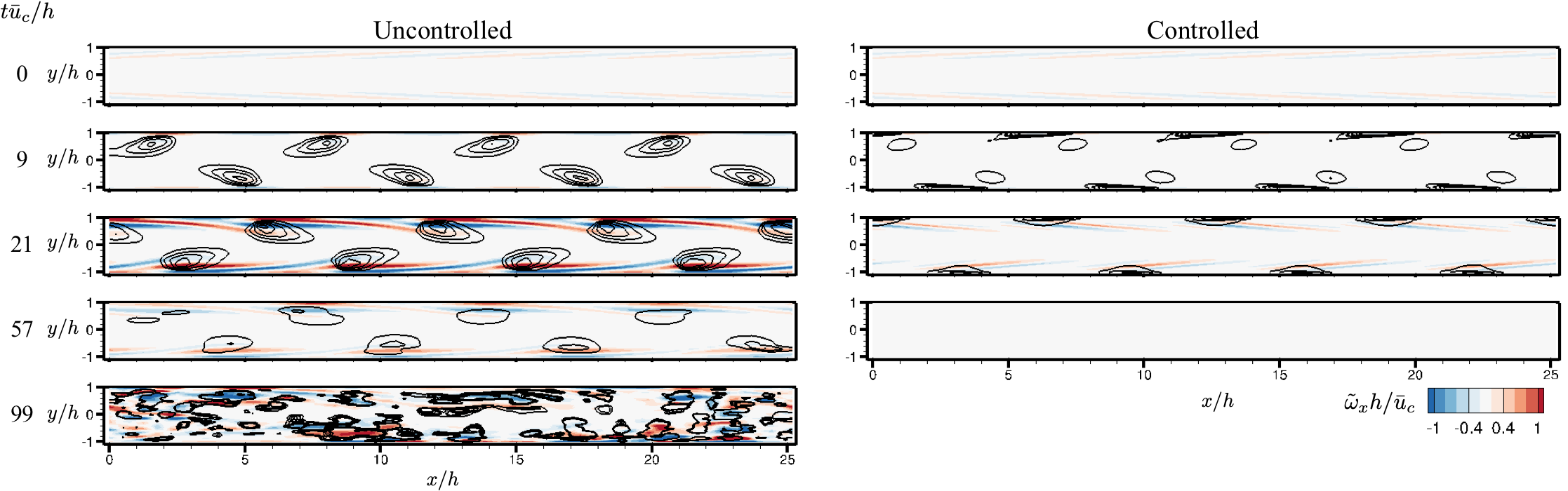}
    \caption{A comparison of instantaneous streamwise vorticity $\tilde \omega_x h/\bar u_c$ flow fields at slice $z/h=0$ between uncontrolled and controlled cases with $(\alpha,\beta)=(1,1)$ and $E_0=0.5\times10^{-4}$. Black contour lines denote $\mathcal Q$-criterion in a range of $0.01 \le \mathcal Q(h/\bar u_c)^2\le 0.05$.}
    \label{fig:slice_A1B1_ctr}
\end{figure}

The transient energy growth present in the uncontrolled flow is caused by a development of large-scale coherent structures, which induce streamwise vortices that further cause a transition to turbulence, as discussed above. A comparison of flow features between uncontrolled and controlled flows during the transition process is shown in Figure \ref{fig:slice_A1B1_ctr}. For a fair comparison, both simulations start from the same optimal disturbance calculated from the uncontrolled flow.  In the uncontrolled flow, large magnitude of the initial optimal disturbance is concentrated near the upper and lower walls where coherent structures denoted by $\mathcal Q$-criterion gradually grow into large-scale coherent structures. When the vertical size of the structures increases to approximately the half height of the channel ($t\bar u_c/h \approx 21$), the kinetic energy of perturbation reaches its maximum value around. During this process, streamwise vortices creep in near the walls following the advection of the coherent structures. After $t\bar u_c/h \approx 21$, the coherent structures start to decay, leaving the induced streamwise vortices in the flow; this is especially evident in the vicinity of the channel walls. As these vortices evolve, break, and interact with each other, the flow becomes chaotic and a laminar-to-turbulent transition occurs. Therefore, the key factor leading to the final transition to turbulence is the breakdown of the streamwise vortices. These streamwise vortices are generated by the rapid growth of the large-scale coherent structures.

\begin{figure}[hbpt]
    \centering
    \includegraphics[width=0.7\textwidth]{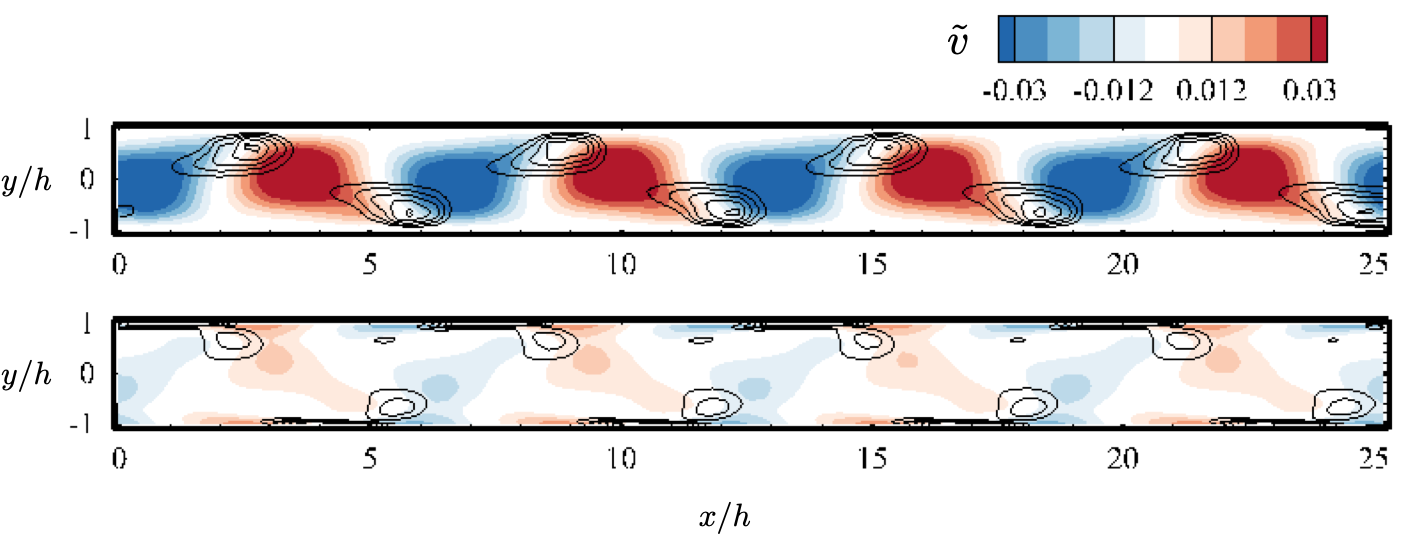}
    \caption{A comparison of instantaneous flow field at slice $z/h=0$ and $tu_c/h=12$ between uncontrolled and controlled flow with $(\alpha,\beta)=(1,1)$ and $E_0=0.5\times10^{-4}$. Contours are wall-normal velocity $\tilde v$, and black contour lines denotes $\mathcal Q$-criterion in a range from 0.01 to 0.05 with increment of 0.01.}
    \label{fig:shear_vbc}
\end{figure}

Compared to the mechanism observed from the uncontrolled flow, the most apparent change in the flow by introducing wall-normal actuation is the growth of the coherent structures. As shown in Figure~\ref{fig:shear_vbc}, the actuation on the wall modifies the distribution of high shear in the flow, in which local areas of high shear stress are created and constrained in the vicinity of the channel walls. As a consequence, the growth of the coherent structure is confined near the walls. Because the strength of these relatively small coherent structures is not enough to induce high-level of streamwise vorticity, the transition observed from the uncontrolled flow is avoided ultimately (Figure~\ref{fig:slice_A1B1_ctr} (right)).

\subsubsection{Controlled flow of spanwise-wave disturbances}

\begin{figure}[hbpt]
    \centering
    \includegraphics[width=0.4\textwidth]{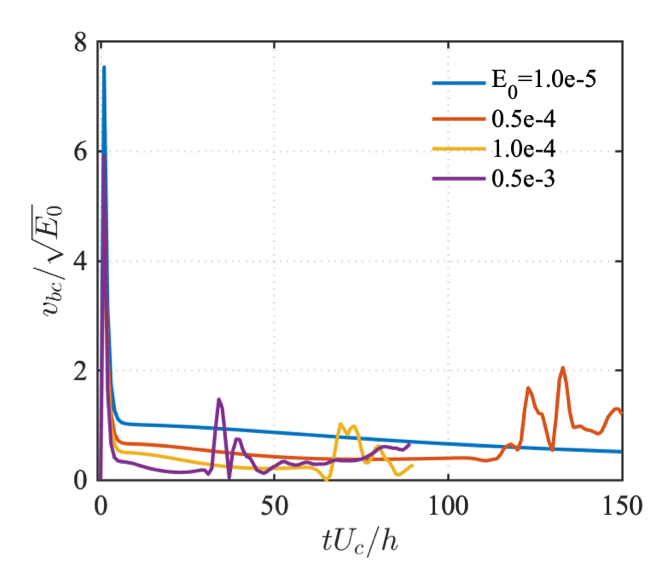}
    \caption{Time history of normalized wall-normal velocity $v_\text{bc}/\sqrt{E_0}$ at lower wall of case $(\alpha,\beta)=(0,2)$ when actuation is turned on. }
    \label{fig:A0B2_vbc}
\end{figure}

 Wall-normal velocity histories at $[x,y,z]/h=[0,-1,0]$ in the controlled flows for spanwise-wave disturbance are shown in Figure \ref{fig:A0B2_vbc}. A sharp increase in wall-normal velocity is induced when the control is turned on, then the velocity quickly decays to a relatively small value and gradually decreases later in time. The oscillations in the wall-normal velocity at later times correspond to the fact that the flow has reached a turbulent state. 

\begin{figure}[hbpt]
    \centering
    \includegraphics[width=0.95\textwidth]{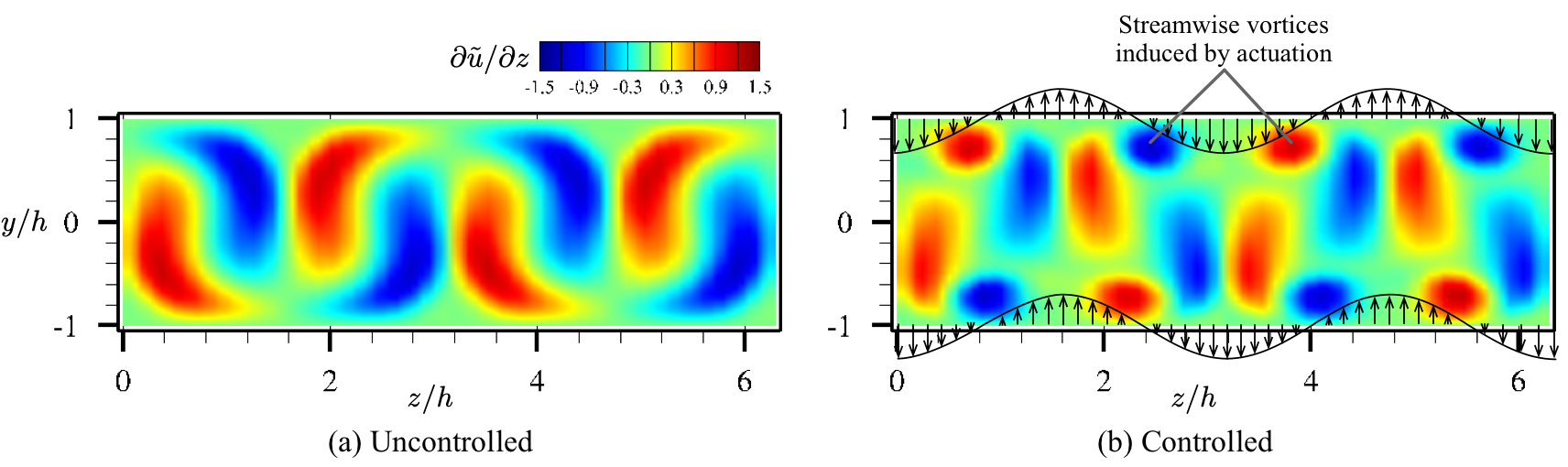}
    \caption{Modification of instantaneous streamwise velocity gradient in spanwise direction ${\partial \tilde u}/{\partial z}$ at slice $x/h=0$ and $t\bar u_c/h=50$ in (b) controlled flow compared to (a) uncontrolled flow. Actuation velocity is denoted by black arrows.}
    \label{fig:slice_A0B2_ctr}
\end{figure}

In the transition process associated with a spanwise-wave disturbance $(\alpha,\beta)=(0,2)$, the merging two co-rotating vortices highly distorts the base velocity profile, as seen in Figure \ref{fig:slice_A0B2_ctr}. The generation of high-shear areas introduces secondary instabilities into the flow for the uncontrolled flow. To examine the control mechanism, we compare the shear flow fields at $t\bar u_c/h=50$ in Figure \ref{fig:slice_A0B2_ctr}. In the uncontrolled flow, high magnitude of shear stress $\partial u/\partial z$ is formed between the merged vortices, where we previously observe instabilities in Figure \ref{fig:dns_base_A0B2_flowfield}. In the controlled flow, a strong actuation forms small-scale streamwise vortices near walls, as seen in Figure \ref{fig:slice_A0B2_ctr} (b). The introduction of these new vortices by control hinders the growth of the merged vortices. Instead, the large streamwise vortical structures are compressed and centralized in the channel. The high shear stress between the vortices is also weakened, as revealed by $\partial \tilde u/\partial z$, providing less chances for secondary instabilities to arise.

\begin{figure}[hbpt]
    \centering
    \includegraphics[width=0.9\textwidth]{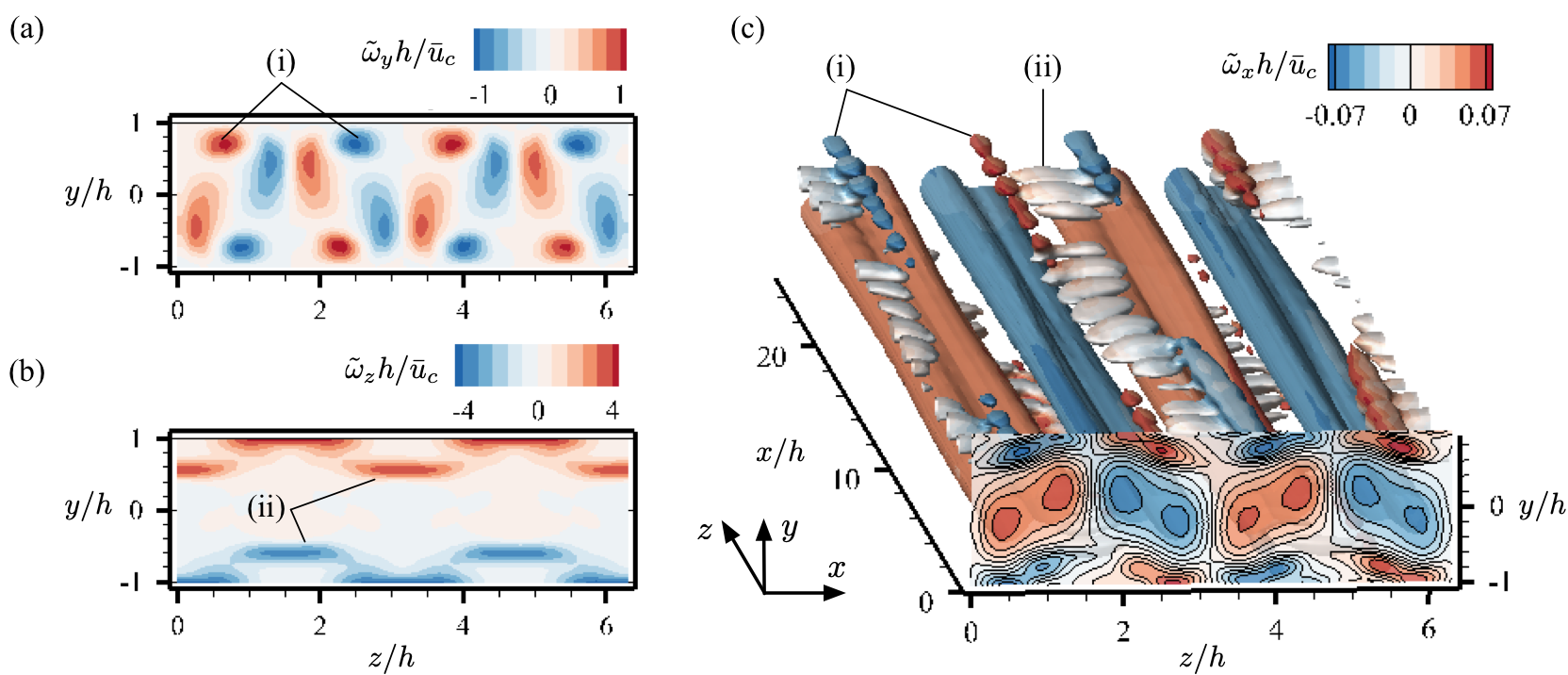}
    \caption{Secondary instabilities induced by the actuation in the flow field at $t\bar u_c/h=66$ of controlled case with $(\alpha,\beta)=(0,2)$ and $E_0=1.0\times 10^{-4}$. (a) Wall-normal vorticity at slice of $x/h=0$, (b) spanwise vorticity at slice of $x/h=0$, and (c) iso-surfaces of $\mathcal Q$ colored by streamwise vorticity. }
    \label{fig:slice_A0B2_ctr_bad}
\end{figure}

Although the controller has suppressed the growth in large streamwise vortical structures, the actuation actually introduces extra high-shear regions. As shown in Figure \ref{fig:slice_A0B2_ctr_bad} (a) and (b), the areas denoted by (i) and (ii) indicate induced high shear stress. Accordingly, as seen in Figure \ref{fig:slice_A0B2_ctr_bad} (c), new secondary instabilities creep in based on the new high-shear regions of (i) and (ii), which ultimately leads to a transition to turbulence. Hence, the secondary instabilities are introduced by the actuation in the nonlinear simulations, which is not accounted for in the controller design from the linear model. 

\subsubsection{Increase in perturbation threshold for laminar-to-turbulent transition}

\begin{figure}[hbpt]
    \centering
    \includegraphics[width=0.65\textwidth]{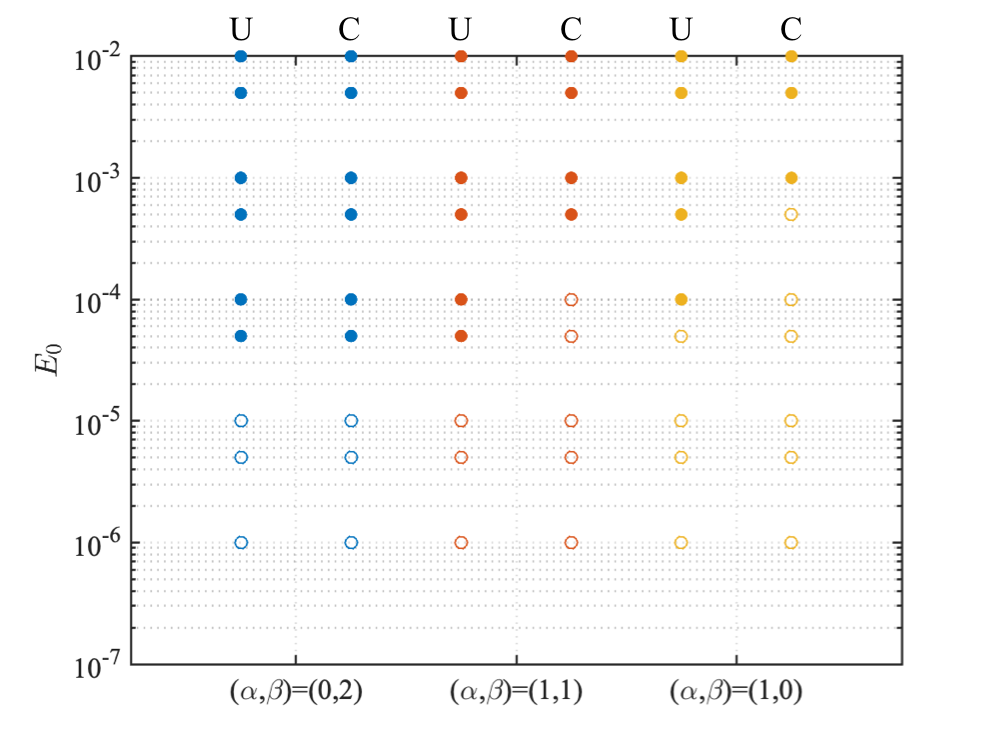}
    \caption{Summary of cases considered in the present work. Solid circle represents that laminar-to-turbulent transition occurs; open circle denotes that the flow remains laminar. U and C denote uncontrolled and controlled cases, respectively.}
    \label{fig:threshold}
\end{figure}

A summary of the cases considered in the present work is provided in Figure \ref{fig:threshold} based on the nonlinear simulation results. For all three wavenumber pairs, the LQR controller successfully suppresses the transient energy growth within linear analysis. In nonlinear simulations, LQR control increases the perturbation threshold $E_0$ for a laminar-to-turbulent transition by roughly an order of $O(10)$ in amplitude in the cases of streamwise-wave $(\alpha,\beta)=(1,0)$ and oblique $(\alpha,\beta)=(1,1)$ disturbances. For the spanwise-wave disturbance $(\alpha,\beta)=(0,2)$, LQR controller successfully suppresses the transient energy growth of the streamwise coherent structures. However, new instabilities are induced by the strong actuation applied on the walls, which leads to laminar-to-turbulent transition by another route; hence, the perturbation threshold does not change based on the cases considered. 

\section{Conclusions}
\label{sec:conclusion}

In this paper, we examined the laminar-to-turbulent transition in plane Poiseuille flow due to transient energy growth of optimal disturbances at a subcritical Reynolds number of $Re=3000$. Nonlinear effects on transient energy growth were investigated by performing direct numerical simulations. We found that with the presence of nonlinearity in the flow, an increase in initial energy density of an optimal disturbance leads to a reduction in amplification rate. Moreover, we uncovered the underlying physics of the transition process for all the disturbances considered. For disturbances associated with $(\alpha,\beta)=(1,1)$ and (1,0), the transient energy growth corresponds to the development of large coherent structures. The evolution of these large-scale coherent structures induces a growth of streamwise vortices near the channel walls, and the breakdown of these vortices leads to a laminar-to-turbulent transition. For a disturbance associated with $(\alpha,\beta)=(0,2)$, the base flow is highly distorted due to the growth of streamwise vortices, which results in high-shear-stress regions formed between the streamwise vortical structures. These high-shear areas introduce secondary instabilities that lead to a laminar-to-turbulent transition.

LQR feedback controllers were designed and shown to be capable of reducing transient energy growth via wall-normal blowing and suction actuation at the upper and lower walls.
  In the controlled flow, the wall actuation modified the shear distribution in the flow, such that coherent structures only formed in the vicinity of the channel walls.
  For streamwise-wave $(\alpha,\beta)=(1,0)$ and oblique $(1,1)$ disturbances, this resulted in a more rapid decay of disturbance energy than for the uncontrolled flow.
Laminar-to-turbulent transition was prevented because the hindered growth of coherent structures significantly inhibits the strength of induced streamwise vortices.
The control strategy was found to increase the threshold for transition by approximately $O(10)$ in magnitude.  For the case with spanwise-wave disturbance of $(\alpha,\beta)=(0,2)$, the controller did not effectively increase the threshold for the transition; the actuation was found to induces extra instabilities that led the flow to transition from a laminar to turbulent state.

The present work has shown the capability of feedback control for suppressing transient energy growth and preventing laminar-to-turbulent transition. The nonlinear effects and underlying physics related to transition process have been investigated by performing direct number simulations. The findings obtained from this study can offer valuable insights into the physical-mechanism that enable transition delay by feedback control.

\section*{Acknowledgments}
This material is based upon work supported by the Air Force Office of Scientific Research under award numbers FA9550-17-1-0252 and FA9550-19-1-0034. We thank Minnesota Supercomputing Institute for providing computational resources. 

\bibliographystyle{aiaa}
\bibliography{ref}

\end{document}